\documentclass[12pt]{article}
\usepackage{geometry}                % See geometry.pdf to learn the layout options. There are lots.
\usepackage[parfill]{parskip}    % Activate to begin paragraphs with an empty line rather than an indent
\usepackage{graphicx}
\usepackage{amsmath, amsthm, amsfonts}
\usepackage{lipsum}
\usepackage{newtxtext,newtxmath}

\newcommand{\Lpmc}[1]{\mathcal{L}^\pm_{#1}}

\usepackage{color}
     \usepackage[colorlinks]{hyperref}
\hypersetup{linkcolor=blue,%
citecolor=red,%
urlcolor=cyan}
\usepackage{url} 
\usepackage{epstopdf}
\title{Factorization Way to Symmetries of
\\
Systems on Curved Spaces
}

\author{ 
 Sergio Salamanca$^z$
\footnote{sergio.salamanca@uva.es, ORCID: 
\href{https://orcid.org/0000-0003-0151-8373}{0000-0003-0151-8373}}   
\bigskip
\\
\noindent
\\ 
 \noindent
$^a$\,Departamento de F\'{\i}sica Te\'orica, At\'omica y
\'Optica, and IMUVA,\\ Universidad de Valladolid,  47011 Valladolid, Spain
}
\begin{document}

\maketitle
\begin{abstract}
In a previous work we showcased the factorization method to find the symmetries of superintegrable systems with spherical separability in flat spaces. Here we analyze the same problem, but in constant curvature spaces along the examples of curved Kepler-Coulomb and Harmonic Oscillator systems. We also show how this procedure can also be directly extended to the curved Smorodinsky-Winternitz (SW) and Evans systems.
\end{abstract}


\noindent
KEYWORDS: superintegrable systems; separation of variables; factorization; intertwining operators; ladder operators; symmetries; action-angle variables, {curvature,integrable system, Harmonic Oscillator, Kepler-Coulomb, lie symmetries, maximally superintegrability.
\section{Introduction}

The aim of this work is to illustrate how the factorization method introduced in \cite{nuestra} can be applied in order to obtain symmetries of  superintegrable systems on constant curvature spaces. The method was originally defined as a tool to analyze separable systems,  a requirement of the method, in order to obtain a symmetry basis in such way that the resulting symmetries allow us to obtain the state spectra and the action-angle variables in the quantum and classical paradigms.
\vspace{5mm}

We are working with central potential systems with spherical coordinates where each separated one-dimensional  problem is connected with the next one through the eigenvalue. In flat space, the sequence of reduced Hamiltonians consists of the $z$-component of the angular momentum, $L^2_z$, the total angular momentum, $L^2$, and the Hamiltonian $H$:
\[
L_z= p_\varphi^2,\qquad
{L}^2= p_{\theta}^2 +\frac{L_z^2}{\sin^2{\theta}}  ,\qquad
H =  p_{r}^2  +V(r) +\frac{L^2}{r^2}
\]
The separated states of the form,
\begin{equation}\label{phi}
\Psi_{n,\ell,m}(r,\theta,\varphi)=
R_n^{\ell}(r)\Phi_{\ell,m}(\theta,\varphi)=
R_n^{\ell}(r)P^m_{\ell}(\theta) \phi_m(\varphi)
\end{equation}
 give rise to three reduced Hamiltonian problems connected by the eigenvalues:
\begin{equation}\label{phi2}
\begin{array}{ll}
\displaystyle 
(a)\quad &L_z= p_\varphi^2\, \phi_m(\varphi) = m^2 \phi_m(\varphi)
\\[2.ex] \displaystyle 
(b)\quad &{L}_m^2 \Phi_{\ell,m}= \big(p_{\theta}^2 +\frac{m^2}{\sin^2{\theta}}\big) \Phi_{\ell,m}
=\ell(\ell+1) \Phi_{\ell,m} 
\\[2.ex] \displaystyle
(c)\quad & H_\ell R_n^{\ell}(r)=  \big(p_{r}^2  +V(r) +\frac{\ell(\ell+1)}{r^2} \big)
R_n^{\ell}(r)= E R_n^{\ell}(r)
\end{array}
\end{equation}

We take advantage of this sequence in order to obtain shift and ladder operators for each one dimensional problem. While ladder operators modify the eigenvalue of a Hamiltonian, shift operators connect Hamiltonians with different parameter values. By coupling a ladder operator of a reduced Hamiltonian with a shift of the next one, we are able to define  operators that will correspond to  symmetries of the system whose effect on the eigenstates is known. The same analysis can be done in the classical paradigm where we obtain constants of motion associated with the action-angle variables.

\vspace{5mm}
Following this procedure, we arrived to the symmetries of systems in flat spaces. 
Here, we considere the application of the method to central potential systems in curved spaces, the generalized Harmonic Oscillator (HO) and Kepler-Coulomb (KC) systems, in order to show the consistency in the flat limit with our previous results \cite{nuestra}
as well as with other standard analysis of these systems \cite{ballesteros,carinena08,carinena12,ballesteros2014new,najafizade2021behavior}. 

Finally, we want to illustrate how the factorization method can be extended to non-central potentials of curved systems which can be separated in spherical coordinates  \cite{evans90,herranz2005maximally,shmavonyan2019cn}. This is done for the  analogs of Smorodinsky-Winternitz (SW) and Evans systems when they are defined on curved spaces where we can compare our results with those obtained in other ways \cite{evans08,winternitz13,ttw09,ttw10}. 

%While these systems are well known, their analysis allow us to illustrate how the application of the factorization method, specially the analysis of systems in curved spaces, can be applied in order to define new supersymmetric systems with known results.

%The analysis of this systems is done in terms of a system identification where we are able to consider the results obtained for the curved HO system 

%The analysis of this systems allow us to illustrate new applications of the factorization method, since the analysis of the angular problems is done in terms of the results obtained for the curved HO. 
\vspace{5mm}

The organization of the work is as follows. In section 2 we introduce the central potential systems on constant curvature spaces. First by the definition of the kinetic energy associated to the geometry of the space and later adding the potentials defined on  the curved HO and KC systems.
In section 3 we construct the ``radial symmetry'' of the curved KC system from its factorization, which in the limit of zero curvature leads to the flat KC version studied in \cite{nuestra}.
Similarly, section 4 contains the analysis of the curved HO system. For completeness, here we have also computed energy ladder operators. We summarize these results and show that in the limit of zero curvature they come into the ones obtained for the flat HO system in \cite{nuestra}.
Section 5 illustrates how to extend the method to non central potential systems by working out the SW and Evans systems.
Section 6 is devoted to the classical symmetries of these systems. In the final section we outline the main advantages of the factorization method.

%%%%%%%%%%%%%%%%%%%%%
\section{Curved Systems}
In this section we introduce the generalization of central potential systems to constant curvature spaces. 
%\vspace{5mm}
The addition of a constant curvature leads to a redefinition of the radial problem \cite{ballesteros}, that leads to hamiltonians of the form
\begin{equation}
   \mathcal{H}= \left(1-\kappa \rho^2\right)p_{\rho}^2+\frac{L^2(\theta,\varphi)}{\rho^2}+V(\kappa,\rho)
\end{equation}
 where $\kappa$ is the curvature parameter.
 While the curvature of the surface determines the kinetic energy, the radial potential must be considered individually. The metric of these systems in spherical coordinates is given by
\begin{equation}
    ds^2=\frac{d\rho^2}{1-\kappa \rho^2}+\rho^2\left(d\theta^2+\sin^2{\theta}d\varphi^2\right)\end{equation}
%\vspace{5mm}

In order to remove the explicit curvature dependence of the system we are going to make use of the following generalized  functions, described in \cite{ballesteros}.
\begin{equation}\label{coord}
\begin{array}{ll}
\mathrm{C}_\kappa(r)= \displaystyle \sum_{l=0}^{\infty}(-\kappa)^{l}\frac{r^{2l}}{(2l)!} 
%=a^-_{\ell+1} a^+_{\ell+1}-\frac{\omega}{2}(2\ell+3) 
=
\left\{ \begin{array}{lll}
\cos{\sqrt{\kappa}r},& \kappa>0;  \\
 1  ,& \kappa=0;    \\
  \cosh{\sqrt{-\kappa}r} , & \kappa<0; & \\
\end{array}\right.
 \\[2.ex]  
 \\[2.ex]
\mathrm{S}_\kappa(r)= \displaystyle 
 \sum_{l=0}^{\infty}(-\kappa)^{l}\frac{r^{2l+1}}{(2l+1)!} 
%=a^-_{\ell+1} a^+_{\ell+1}-\frac{\omega}{2}(2\ell+3) 
=
\left\{ \begin{array}{lll}
 \frac{1}{\sqrt{\kappa}} \sin{\sqrt{\kappa}r},& \kappa>0;  \\
 r  ,& \kappa=0;    \\
  \frac{1}{\sqrt{-\kappa}} \sinh{\sqrt{-\kappa}r} , & \kappa<0; & \\
\end{array}\right.
 \\[1.ex]  
 \end{array}
\end{equation}

They satisfy the a normalization equation:
$\mathrm{C}^2_{\kappa}(r)+\kappa\mathrm{S}^2_{\kappa}(r)=1
$.
\vspace{5mm}
In the same spirit the $\kappa$-tangent function is defined by
$\mathrm{T}_{\kappa}(r)=\mathrm{S}_{\kappa}(r)/\mathrm{C}_{\kappa}(r)$.

Next, we perform a  change of canonical variables:
\begin{equation}
    \rho=\mathrm{S}_{\kappa}(r) \qquad p_{\rho}=\frac{p_r}{\mathrm{C}_{\kappa}(r)}
\end{equation}
This will allow us to express the Hamiltonian of the system as
\begin{equation}\label{Htrigo}
 \mathcal{H}= p^2_r+\frac{L^2(\theta,\varphi)}{\mathrm{S}^2_{\kappa}(r)}+V(r,\kappa)
\end{equation}
Now, the curvature dependence is hidden inside these new trigonometric functions.
Since the curvature affects only the radial variable, we can make use of the results presented in \cite{nuestra} and omit the angular analysis. 
The separated solutions have the same form than in the flat case (\ref{phi}).
%\begin{equation}\label{phi}
%\Psi_{n,\ell,m}(r,\theta,\varphi)=
%R_n^{\ell}(r)\Phi_{\ell,m}(\theta,\varphi)=
%R_n^{\ell}(r)P^m_{\ell}(\theta) \phi_m(\varphi)
%\end{equation}
The stationary equation in the quantum formalism is: 
\begin{equation}{\label{hamiltonian}}
\mathcal{H}\Psi_{n,\ell,m}(r,\theta,\varphi)=\left(-\partial_{rr}-\frac{2}{\mathrm{T}_{\kappa}(r)}\partial_{r}+\frac{L^2(\theta,\varphi)}{\mathrm{S}^2_{\kappa}(r)}+V(r,\kappa)\right) \Psi_{n,\ell,m}(r,\theta,\varphi)=E_{n}\Psi_{n,\ell,m}(r,\theta,\varphi)
\end{equation}
Regarding the systems separability, only the radial Hamiltonian differs from the flat case (\ref{phi2}), that now takes the form
\begin{equation}\label{hradial}
\mathcal{H}_{\ell}R_n^{\ell}(r)=\left(-\partial_{rr}-\frac{2}{\mathrm{T}_{\kappa}(r)}\partial_{r}+\frac{\ell(\ell+1)}{\mathrm{S}^2_{\kappa}(r)}+V(r,\kappa)\right) R_n^{\ell}(r)=E^\ell_{n}R_n^{\ell}(r)
\end{equation}
Next we are going to apply the factorization method to the curved Harmonic Oscillator and Kepler-Coulomb whose potentials take the following form
\cite{ballesteros,enciso} : 
\begin{equation}\label{pot}
   \mathcal{V_{KC}}=-\frac{q}{\mathrm{T}_{\kappa}(r)},
   \qquad 
   \mathcal{V_{HO}}=\frac{\Omega^2}{4}\mathrm{T}^2_{\kappa}(r)=\frac{\omega^2-k^2}{4}\mathrm{T}^2_{\kappa}(r),
   \qquad   q,\Omega>0,\omega> |k|
   \end{equation} 
The curvature dependence is also hidden inside the generalized trigonometric functions. 
%   However we are going to consider both expressions of the $HO$ potential as they will be proven useful.
%   \newpage
In the plots of effective potentials (\ref{hradial}), we can appreciate how the curvature  mainly modifies the behaviour of the effective radial potentials at the (right) asymptotic regions. The KC effective potential
\[
\mathcal{V_{KC}}^{\rm eff}(r)=-\frac{q}{\mathrm{T}_{\kappa}(r)}+\frac{\ell(\ell+1)}{\mathrm{S}^2_{\kappa}(r)}
\] 
has a behaviour that depends on the curvature value:
\[
{\rm for}\ \kappa>0,\quad \mathcal{V}_{\rm KC}^{\rm eff}(r) \stackrel{ r \to \pi/\sqrt{k}}{\longrightarrow} \infty,\qquad
{\rm for}\ \kappa=0,\quad \mathcal{V}_{\rm KC}^{\rm eff}(r) \stackrel{ r \to \infty}{\longrightarrow} 0,\quad\qquad
{\rm for}\ \kappa<0,\quad \mathcal{V}_{\rm KC}^{\rm eff}(r) \stackrel{ r \to \infty}{\longrightarrow}
-q
\]
In the case of the HO, the effective potential
\[
\mathcal{V_{HO}}^{\rm eff}(r)=\frac{\Omega^2}{4}\mathrm{T}^2_{\kappa}(r)+\frac{\ell(\ell+1)}{\mathrm{S}^2_{\kappa}(r)}
\] 
has the following limits:
\[
{\rm for}\ \kappa>0,\ \mathcal{V}_{\rm HO}^{\rm eff}(r) \stackrel{ r \to\pi/\sqrt{k}}{\longrightarrow} \infty,\qquad
{\rm for}\ \kappa=0,\ \mathcal{V}_{\rm HO}^{\rm eff}(r) \stackrel{ r \to \infty}{\longrightarrow} \infty,\quad,\qquad
{\rm for}\ \kappa<0,\ \mathcal{V}_{\rm HO}^{\rm eff}(r) \stackrel{ r \to \infty}{\longrightarrow}
\frac{\Omega^2}{4}
\]
%However, this effect is more pronounced in the HO system where negative curvature values lead to constant limit leading to a maximum  bound state energy, limit that does not exist for other curvature values.  

\begin{figure}[h]\label{potim}
	\centering
\includegraphics[width= 15 cm]{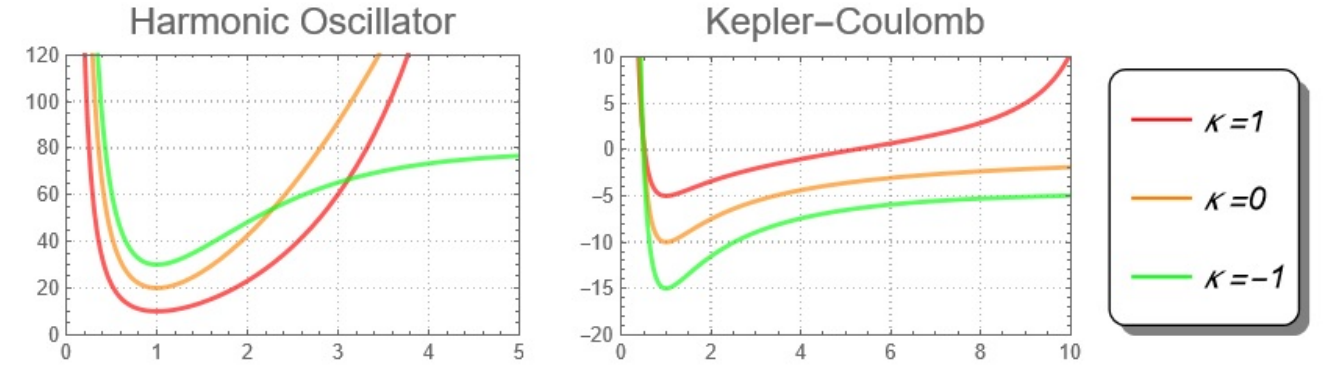}
\end{figure}

In the next two sections we are going to detail the application of the factorization method to the obove quantum KC and HO systems and showcase the results.

%%%%%%%%%%%%%%%%%%%%%%%%%%%%%%

We will start by the curved KC system. The program is as follows: 

(i) There are three commuting symmetries: $L_z^2$, $L^2$, and $H$. 

(ii)
Then, we will compute two additional pairs of symmetries. The first one is the angular momentum operators ${\cal L}^\pm$ which are  well known.
They are the same for any value of the curvature $\kappa$.

(iii) The second pair of symmetries is obtained (as shown in ref[nuestra}])
as the product of a shift operator of the reduced radial Hamiltonian
$\mathcal{H}_{\ell}$ and
a ladder operator, $\Lambda_\ell$ of the reduced angular Hamiltonian $L_m^2(\theta)$.

The only new symmetry we must calculate is that of (iii). We already have one of the ingredients: the ladder operator $\Lambda^\pm_\ell$, because it coincides with that one of the flat system cite[nuestra]. Its action is to change the eigenvalue parameter $\ell$ of $L_m^2$,
\begin{equation}\label{lad}
L^2_m \Phi_{\ell,m} = \ell(\ell+1),\qquad \Lambda^-_\ell \Phi_{\ell,m} \propto \Phi_{\ell- 1,m},\quad 
\Lambda^+_{\ell+1} \Phi_{\ell,m} \propto \Phi_{\ell+ 1,m}
\end{equation}
%These ladder operators were already computed in [nuestra].

\section{Curved Kepler-Coulomb}
We only need to compute the shift operator of the reduced radial Hamiltonian
\begin{equation}\label{hradial2}
\mathcal{H}^{\rm KC}_{\ell} %R_n^{\ell}(r)
=\left(-\partial_{rr}-\frac{2}{\mathrm{T}_{\kappa}(r)}\partial_{r}+\frac{\ell(\ell+1)}{\mathrm{S}^2_{\kappa}(r)} -\frac{q}{\mathrm{T}_{\kappa}(r)}\right)  %R_n^{\ell}(r)=E_{n}R_n^{\ell}(r)
\end{equation}

This is obtained from the standard factorization of the effective Hamiltonian (\ref{hradial2}):

\begin{equation}\label{kcurvo}
\mathcal{H}^{\rm KC}_\ell=\Sigma^+_{r, \ell} \Sigma^-_{r, \ell}-\frac{q^2}{4\ell^2}+\kappa(\ell^2-1)=\Sigma^-_{r, \ell+1} \Sigma^+_{r, \ell+1}-\frac{q^2}{4(\ell+1)^2}+\kappa\ell(\ell+2),
\quad
 \left\{ \begin{array}{ll}
   \Sigma^+_{r, \ell}=-\partial_r+\frac{\ell-1}{\mathrm{T}_{\kappa}(r)}-\frac{q}{2\ell},  &   \\
     &  \\[1.5ex]
    \Sigma^-_{r, \ell}=\partial_r+\frac{\ell+1}{\mathrm{T}_{\kappa}(r)}-\frac{q}{2\ell},   
    &  \\
\end{array}\right.
\end{equation}

Due to their definition, these factorization operators $\Sigma^\pm_{r, \ell}$  supply us with key properties of the system: 
%\vspace{5mm}

On one hand, from the factorization equation we can identify the shift character of this operators, as they connect Hamiltonians and eigenfunctions with different $\ell$-values while maintaining energy:
\begin{equation}\label{shiftkc}
\begin{array}{lll}
\displaystyle  \Sigma^+_{r, \ell+1}\mathcal{H}^{\rm KC}_{\ell}=\mathcal{H}^{\rm KC}_{\ell+1}\Sigma^+_{r, \ell+1} \quad &\rightarrow\quad   &\Sigma^+_{r, \ell+1}R_n^{\ell}(r)\propto R_n^{\ell+1}(r)     
\\[2ex]
 \displaystyle   \Sigma^-_{r, \ell}\mathcal{H}^{\rm KC}_{\ell}=\mathcal{H}^{\rm KC}_{\ell-1}\Sigma^-_{r, \ell} &\rightarrow   & \Sigma^-_{r, \ell}R_n^{\ell}(r)\propto R_n^{\ell-1}(r)  
\end{array}
\end{equation}

On the other  hand, by definition the states annihilated by the factorization operators satisfy the Hamiltonian equation, in this case, they annihilate radial states with maximum angular momentum value which have associated the energy of the system:
\begin{equation}\label{kcshiftest}
    \Sigma^+_{r, n+1}R_n^n(r)=0\ \rightarrow \  R_n^n(r)= e^{\frac{-q r}{2(n+1)}}\mathrm{S}^{n}_{\kappa}(r)
\end{equation}
\begin{equation}
  \mathcal{H}^{\rm KC}_n R_n^n(r)= E_{n} R_n^n(r) \ \implies  \  E_{n}=\kappa n(n+2)-\frac{q^2}{4(n+1)^2} 
\end{equation}
Where the energy $E_n$ of the curved KC system contains a new  $\kappa$--dependent term corresponding to the energy associated to the angular momentum value in curved spaces, plus the standard energy term of flat KC system.

\vspace{5mm}

We can see that these operators are  continuous in the curvature and they constitute a generalization of the shift operators for the flat KC system  which are obtained in the limit of zero curvature \cite{nuestra}:  
\begin{equation}
    \lim_{\kappa\rightarrow 0}{}_{r}\Sigma_{\ell} ^\pm= \mp \partial_r+\frac{\ell\mp1}{r}-\frac{q}{2\ell}
\end{equation}

%\newpage

 Once obtained, the shift operators, we may construct the symmetries
 ${}_{r,\theta}\mathcal{S}^\pm_{\ell}$, following the structure of the flat problem, by means of the product of the  radial shift operators
 ${}_{r}\Sigma_{\ell} ^\pm$, with 
 the $\theta$-ladder operators  ${}_{\theta}\Lambda_{\ell}^\pm$ shown in (\ref{lad}),
% that connect $\theta$-states $P^m_{\ell}(\theta)$ with different $\ell$:
 \begin{equation}\label{laddang}
  {}_{r,\theta}\mathcal{S}^-_{\ell}={}_{r}\Sigma_{\ell}^-\, {}_{\theta}\Lambda_{\ell}^-;
  \qquad
  {}_{r,\theta}\mathcal{S}^+_{\ell+1}={}_{r}\Sigma_{\ell+1}^+\, {}_{\theta}\Lambda_{\ell+1}^+  
%  \quad {}_{r,\theta}\mathcal{S}^+_{\ell+1}\Psi_{n,\ell,m}\propto\Psi_{n,\ell+1,m};\quad {}_{r,\theta}\mathcal{S}^+_{\ell}\Psi_{n,\ell,m}\propto\Psi_{n,\ell-1,m} 
\end{equation}
These symmetry operators match their action on the components of separated eigenfunctions and connect states with different angular momentum values while preserving the  energy:
\begin{equation}\label{laddang}
%  {}_{r,\theta}\mathcal{S}^\pm_{\ell}={}_{r}\Sigma_{\ell+1} ^\pm {}_{\theta}\Lambda_{\ell}^\pm;  \quad 
  {}_{r,\theta}\mathcal{S}^+_{\ell+1}\Psi_{n,\ell,m}\propto\Psi_{n,\ell+1,m};\qquad 
  {}_{r,\theta}\mathcal{S}^-_{\ell}\Psi_{n,\ell,m}\propto\Psi_{n,\ell-1,m} 
\end{equation}

 The combined action of the above symmetries $\mathcal{S}^\pm_{\ell}$ with the angular symmetries ${\cal L}^\pm$, allow us to obtain the  state basis of a degeneracy subspace from a highest weight eigenfunction:
\begin{equation}\label{estadoskc}
\mathcal{S}^+_{n+1}\Psi_{n,n,n}%=L^+\Psi_{n,n,n}
=0\ \implies\   \Psi_{n,n,n}\propto e^{\frac{-q r}{2(n+1)}+i n\varphi}\mathrm{S}^{n}_{\kappa}(r)\sin^{n}{\theta} \ ;
 \qquad
 \Psi_{n,\ell,m}\propto \left(L^-\right)^{n-m} \prod_{i=0}^{n-\ell}\mathcal{S}^-_{n-i}\Psi_{n,n,n}
\end{equation}
 However, the eigenfunctions $\Psi_{n,\ell,m}$ must be square integrable in order to represent physical states. This leads to a energy restriction for $\kappa<0$ as we consider the integrability of the highest weight states:
\begin{equation}\label{nmax}
    \int |\Psi_{n,n,n}|^2 dV =1\rightarrow 2(n+1)^2\sqrt{-\kappa}<q\quad \kappa<0
\end{equation}
This energy restriction can be interpreted as an imposition of increasing energy values:
\begin{equation}
 \frac{\partial E_n}{\partial n}=   2\kappa (n+1)+\frac{q^2}{2(n+1)^3}<0
\end{equation}
 While this equation is always satisfied for $\kappa\geq 0$, it leads to (\ref{nmax}) for negative curvature values. 
Taking into account this restriction on $n$ for $\kappa<0$, 
the allowed $n,\ell,m$ parameters must satisfy the following 
inequalities in order to represent physical states:
\begin{equation}\label{parametroskc}
   0\leq |m| \leq \ell \leq n  ;\qquad 2\kappa (n+1)+\frac{q^2}{2(n+1)^3}<0 ;\qquad m,\ell,n \in \mathbb{N} 
\end{equation}

%%%%%%%%%%%%%%%%%%%%%%%%%%%%%
 \section{Curved Harmonic Oscillator}
 
 In this section we will present a schematic analysis of the curved HO system following the line of the previous section for  curved KC system,  a deeper rundown of this section is contained in the appendix.
 
 \vspace{5mm}
We are working with the following reduced radial equation (\ref{hradial}):
\begin{equation}{\label{hradialtan}}
\mathcal{H}^{\rm HO}_{\ell,\omega}R_n^{\ell}(r)=\left(-\partial_{rr}-\frac{2}{\mathrm{T}_{\kappa}(r)}\partial_{r}+\frac{\ell(\ell+1)}{\mathrm{S}^2_{\kappa}(r)}+\frac{\omega^2-\kappa^2}{4}\mathrm{T}^2_{\kappa}(r)\right) R_n^{\ell}(r)=E_{n}R_n^{\ell}(r) \end{equation}
Notice that this is a type of P\"oschl-Teller Hamiltonian. Its factorization properties are well known; they have some differences from those of KC. In fact, there are  two pairs of factorization operators $a^\pm$ and $b^\pm$. Similarly to the factorization of the KC system, 
 these factorization operators connect reduced HO Hamiltonians but this time with two different parameters: $\ell$ and $\omega$. 
% in this case modifying both the $\ell$ and $\omega$ parameters
 At the same time, they induce  an energy change (we have omitted  subindexes of
  $a^\pm,b^\pm$ for simplicity, more details are given in the appendix):
\begin{equation}\label{commut0}
    \begin{array}{ll}
a^\pm%_{\ell+1,\omega-2\kappa} 
\mathcal{H}^{\rm HO}_{\ell,\omega}=\left(\mathcal{H}^{\rm HO}_{\ell\pm 1,\omega \mp 2k}+(\kappa\mp\omega)\right)a^+%_{\ell+1,\omega-2\kappa}
, \qquad
%  a^-_{\ell,\omega} 
%  \mathcal{H}_{\ell,\omega}=\left(\mathcal{H}_{\ell-1,\omega+2k}+(\kappa+\omega)\right)a^-_{\ell,\omega}
   \\[2.ex]
     b^\pm%_{\ell\pm 1,\omega\pm 2\kappa}
     \mathcal{H}^{\rm HO}_{\ell,\omega}=\left(\mathcal{H}^{\rm HO}_{\ell\pm 1,\omega\pm 2k}+(\kappa\pm\omega)\right)b^\pm%_{\ell+1,\omega+2\kappa}
%     , \qquad
%  b^-_{\ell,\omega} \mathcal{H}_{\ell,\omega}=\left(\mathcal{H}_{\ell-1,\omega-2k}+(\kappa-\omega)\right)b^-_{\ell,\omega}
\end{array}
\end{equation}

  The effect of these operators on the parameters of the HO radial effective hamiltonian, $\mathcal{H}^{\rm HO}_{\ell,\omega}$, is summarized in the following graph:
\begin{figure}[h]
	\centering
\includegraphics[width=7 cm]{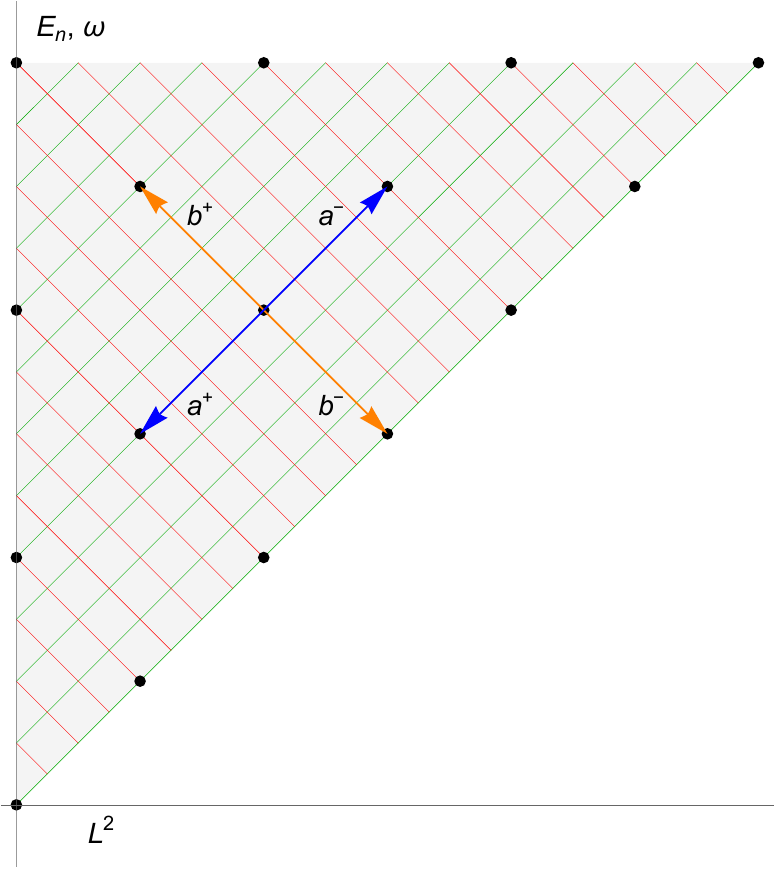}
\caption{Schematic representation of the action of the auxiliary operators $a^\pm(r)$ and $b^\pm(r)$ where each dot $(\ell,n,\omega)$ represent an eigenfunction   $R_n^\ell(r)$ of $\mathcal{H}^{\rm HO}_{\ell,\omega}$.
\label{figx}}
\end{figure}

Next,  we are able to get true shift operators ${}_{r}\Sigma ^\pm$
%isolate the shift effect allowing us to define operators 
that modify only the $\ell$-parameter in two units, while preserving the energy, by means of the products:
% based on the effect of $a^\pm$ and $b^\pm$. We are able to obtain symmetry operators as the coupling of this radial shift, of the form 
 ${}_{r}\Sigma_\ell ^\pm=a^\pm b^\pm$. 
 
Then, in order to build the symmetries we should match the shift operator 
${}_{r}\Sigma ^\pm$ with a ladder operator in $(\theta,\ell)$  changing  
the parameter $\ell$ also in two units, in the same spirit as (\ref{laddang}). This is done by applying two times the operator ${}_\theta\Lambda_\ell$ given in (\ref{lad}):
%ladder operator. Matching the $\ell$ change rate in both coordinates we arrive to the following symmetry operators:
\begin{equation}
   {}_{r,\theta}\mathcal{S}^\pm={}_{r}\Sigma_\ell^\pm \left({}_{\theta}\Lambda_\ell^\pm\right)^2; 
  \qquad {}_{r,\theta}\mathcal{S}^\pm\Psi_{n,\ell,m}(r,\theta,\varphi)\propto\Psi_{n,\ell\pm 2,m}
\end{equation}
Similarly to the KC system, the symmetries ${}_{r,\theta}\mathcal{S}^\pm$ obtained are continuous in the curvature and can be seen as  generalizations of the ones obtained in \cite{nuestra}. Furthermore we are able to obtain the eigenvalue   spectrum and eigenstates of the system from the factorization of
$\mathcal{H}^{\rm HO}_{\ell,\omega}$ in terms of  $a^\pm, b^\pm$:
\[
\mathcal{H}^{\rm HO}_{\ell,\omega}\, \Psi_{n,\ell,m}=  E_\ell^{n}\Psi_{n,\ell,m}; \qquad E_\ell^n=\ \kappa(n(n+3) +\omega(n+\frac{3}{2}),\qquad \ell\leq n 
\] 
\bigskip

A key difference in the analysis of the KC and HO systems is that while we are able to obtain independent symmetry operators for both systems, the HO admits energy ladder operators. The construction is detailed in the appendix. Here we mention that in the same way we used above, there are two pairs of operators, $c^\pm, d^\pm$, which shift the parameters $\omega, n$ corresponding to  frequency and energy of an eigenfunction
of $\mathcal{H}^{\rm HO}_{\ell,\omega}$. The action of these operators is to change such parameters $(\omega,n)$ in $\pm(2\kappa,1)$ and
$\pm(2\kappa,-1)$. Then the pure ladder operators ${}_{r}\Lambda^\pm$ changing only the energy have the form
$c^+d^+$ and $c^-d^-$,
%These factorization operators are continuous in the curvature, and while they do not have a flat equivalent, they can be coupled in order to obtain energy ladder operators, that correspond to a generalization of the energy ladder operators obtained in \cite{nuestra} : 
\begin{equation}
    {}_{r}\Lambda^\pm= \frac{1}{\kappa
}d^\pm c^\pm; \quad  {}_{r}\Lambda^\pm\Psi_{n,\ell,m}\propto \Psi_{n\pm 2 ,\ell,m}
\end{equation}

%\vspace{5mm}

%In order to obtain this ladder operators we define an auxiliary operator $\mathrm{N}_{n,\omega}$ based on the state equation (\ref{hradialtan}) where we consider the energy as a parameter of the problem and $\ell$ as the eigenvalue of the problem. Since this operator represents the same state equation as the Hamiltonian, if we are able to obtain operators that connect $\mathrm{N}_{n,\omega}$ systems with different n values, they will translate into energy ladder operators.
%Applying the factorization method to  $\mathrm{N}_{n,\omega}$ leads to two pair of factorization operators, $c^\pm,d^\pm$. These operators modify both the energy parameter $n$ as well as the frequency of the $\mathrm{N}_{n,\omega}$ operator. The effect of these operators is contained in the following graph:
%
%\begin{figure}[h]
%	\centering
%\includegraphics[width=7 cm]{graficocurvoholddernoindex.pdf}
%\caption{Schematic representation of the action of the auxiliary operators $c^\pm(r)$ and $d^\pm(r)$ where each dot $(n,\omega)$ represent a state $R_n^\ell(r)$ based on their effect on $\mathrm{N}_{n,\omega}$.
%\label{figlad}}
%\end{figure}

In summary, we have obtained operators that modify each of the parameters of the eigenfunction components independently:
\begin{equation}\label{estadosho}
{}_{r}\Lambda^\pm\Psi_{n,\ell,m}\propto\Psi_{n \pm 2,\ell,m};\quad {}_{r,\theta}\mathcal{S}^\pm\Psi_{n,\ell,m}\propto\Psi_{n,\ell\pm 2,m};\quad 
 \mathcal{L}^\pm\Psi_{n,\ell,m}(r,\theta,\varphi)\propto\Psi_{n,\ell,m\pm 1}
\end{equation}
These operators allow us to obtain the state spectrum taking into account their annihilation properties, see appendix. The allowed parameter values are obtained considering the restriction of square integrable states:
\begin{equation}\label{parametroskc}
   0\leq |m| \leq \ell \leq n  ;\quad k (3 + 2 n) + \omega>0 ;\quad m,\ell,n \in \mathbb{N} 
\end{equation}
Where the energy restriction can be obtained by considering increasing energy values. Similarly to the KC system, this energy restriction is only relevant for negative curvature spaces where the curvature diminishes the system capability of state capture.

%Where the associated energy of the states corresponds to:
%\begin{equation}
% \mathcal{H} \Psi_{n,\ell,m}=  E_{n}\Psi_{n,\ell,m}; \quad  E_{n}= \kappa n (n + 2) +(\omega+\kappa)(n+\frac{3}{2}) 
%\end{equation} 

\section{Smorodinsky-Winternitz and Evans Systems}
In this section we undertake an analysis of the Smorodinsky-Winternitz and Evans systems in terms of the factorization method.  
\vspace{3mm}
The potentials of such systems can be interpreted as modifications of the central potential systems, HO and KC by means of  centrifugal terms $V_c$, respectively,
\begin{equation}\label{swep}
    V_{SW}=\frac{\Omega^2}{4}\mathrm{T}^2_\kappa(r) +V_{c},
    \qquad     V_{E}=-\frac{q}{\mathrm{T}_\kappa(r)}+V_{c}
\end{equation}
where the additional centrifugal potential, $V_c$  has the form the form:
\begin{equation}\label{swecp}
    V_{c}
    %=\frac{K_{1}^2}{x^2}+\frac{K_{2}^2}{y^2}+\frac{K_{3}^2}{z^2}
    =\frac{1}{\mathrm{S}^2_\kappa(r)}\left(\frac{k_{1}(k_{1}-1)}{\cos^2{\theta}}+\frac{1}{\sin^2{\theta}}\left(\frac{k_{2}(k_{2}-1)}{\sin{\varphi}^2}+\frac{k_{1}(k_{1}-1)}{\cos{\varphi}^2}\right)\right); \quad k_{i}>0
\end{equation}
This structure leads to a total potential $\tilde V(\varphi,\theta,r)$ compatible with the separability of the Hamiltonian in spherical coordinates including three one-dimensional potentials $\tilde V_r(r), \tilde V_{\theta}, \tilde V_{\varphi}$, of the form: 
\begin{equation}\label{potang}
\tilde V(\varphi,\theta,r)= \tilde V_r(r)+ \frac{1}{r^2}\left(\tilde V_{\theta}(\theta)+\frac{\tilde V_{\varphi}(\varphi)}{\sin^2{\theta}}\right)
\end{equation}
Then, the original three-dimensional problem 
%considering states (\ref{phi}) 
leads to a three one-dimensional sequence of reduced Hamiltonians:
\begin{equation}\label{reduced}
\begin{array}{ll}
(a)\quad &\displaystyle \tilde{\mathcal{L}}_z^2(\varphi) \phi_m(\varphi) := 
\left(-\partial_{\varphi\varphi}+\frac{k_2(k_2-1)}{\sin{\varphi}^2}+\frac{k_1(k_1-1)}{\cos{\varphi}^2}\right)\phi_{\tilde{m}}(\varphi)=\tilde{m}^2 \phi_{\tilde{m}}(\varphi)
\\[2.ex]
(b)\quad &\displaystyle  \tilde{\mathcal{L}}_{\tilde{m}}^2(\theta)P_\ell^{\tilde{m}}(\theta):= 
\left(-\partial_{\theta\theta}-\frac{1}{\tan{\theta}}\partial_\theta
+\frac{\tilde{m}^2}{\sin^2{\theta}}+\frac{k_3(k_3-1)}{\cos{\theta}^2}\, \right)
P_\ell^m(\theta) 
= \tilde\ell(\tilde\ell+1)\, P_\ell^m(\theta)
\\[2.ex]
(c)\quad &\displaystyle \tilde{\mathcal{H}}_{\tilde{\ell},\omega}R_n^{\tilde{\ell}}(r)=\left(-\partial_{rr}-\frac{2}{\mathrm{T}_{\kappa}(r)}\partial_{r}
+\frac{\tilde{\ell}(\tilde{\ell}+1)}{\mathrm{S}^2_{\kappa}(r)}+ V(r)\right) R_n^{\tilde{\ell}}(r)=E_{n}R_n^{\tilde{\ell}}
\end{array}
\end{equation}
Each reduced Hamiltonian is of P\"oschl-Teller type, where their shift and ladder operators are well known [Sengul12]. 

The addition of the centrifugal terms does not affect to the radial problem, therefore we can recover the results of the central potential analysis of the HO and KC radial problems of the previous sections, in particular  the shift and ladder radial operators, described in \cite{nuestra}. 

Therefore, we can construct their symmetries in a straighforward way. Next, we will enumerate the consequences of the factorization method.

\begin{enumerate}

\item
Each reduced Hamiltonian (c) and (b) has second order shift operators 
${}_r\Sigma^\pm_\ell$ and ${}_\theta\Sigma^\pm_\ell$.
\[
{}_r\Sigma^\pm_\ell\, R_n^{\tilde{\ell}} \propto R_n^{\tilde{\ell}\pm2}\,,
\qquad
{}_\theta\Sigma^\pm_\ell\,P_\ell^{\tilde{m}} \propto P_\ell^{\tilde{m}\pm2}
\]

\item
Each reduced Hamiltonian (b) and (a) has second order ladder operators 
${}_\theta\Lambda^\pm_\ell$ and ${}_\varphi\Lambda^\pm_m$.
\[
{}_\theta\Lambda^\pm_{\tilde{\ell}}\,P_{\tilde{\ell}}^{\tilde{m}} 
\propto P_{\tilde{\ell}\pm2}^{\tilde{m}}
\qquad
{}_\varphi\Lambda^\pm_m \,\phi_{\tilde{m}} \propto \phi_{\tilde{m}\pm2}
\]

\item 
There are two symmetries of fourth order:
\[
{}_r{\cal S}_\theta^\pm = {}_r\Sigma^\pm_\ell\,{}_\theta\Lambda^\pm_\ell,\qquad
{}_\theta{\cal S}_\varphi^\pm = {}_\theta\Sigma^\pm_m\,{}_\varphi\Lambda^\pm_m
\]
With associated action,
\begin{equation}
{}_{\theta\varphi}\mathcal{S}^\pm\Psi_{n,\ell,m}\propto\Psi_{n,\ell,m\pm 2}
\quad 
     {}_{r,\theta}\mathcal{S}^\pm\Psi_{n,\ell,m}\propto\Psi_{n,\ell\pm 2,m}
\end{equation}

\item The eigenvalues have the same formulas as shown in (\ref{reduced}) but with different range of values:
\[
\tilde m= (k_1+k_2 +2p),\qquad
\tilde \ell= (k_3+m+2q)= (k_1+k_2+k_3+2q),\qquad,\qquad p,q = 0,1,2,\dots
\tilde n = n
\]
The inequalities  of the values  $\tilde m,\tilde \ell, \tilde n$ is the same as 
for the corresponding central systems:
\[
\tilde n\geq \tilde\ell \geq \tilde m+ k_3\geq k_1+k_2
\] 
The highest weight eigenfunctions and eigenvalues are obtained by considering the annihilation properties of the ladder symmetries, 
%leading to the following bound states:
\begin{equation}
    \begin{array}{cc}
  \Psi_{\Sigma k_{i},\Sigma k_{i},k_{1}+k_{2}}=  \mathrm{C}^{\frac{\kappa+ \omega}{2\kappa}}_{\kappa}(r)
  \mathrm{S}^{\Sigma k_{i}}_{\kappa}(r)\left(\sin{\theta}\right )^{k_{1}+k_{2}} \cos^{k_{3}}{\theta}\cos^{k_{1}}{\varphi}\sin^{k_{2}}{\varphi} &  
  {\rm SW}\\[1.5ex]
  \Psi_{\Sigma k_{i},\Sigma k_{i},k_{1}+k_{2}}= e^{\frac{-q r}{2(n+1)}} 
  \mathrm{S}^{\Sigma k_{i}}_{\kappa}(r)\left(\sin{\theta}\right )^{k_{1}+k_{2}} \cos^{k_{3}}{\theta}\cos^{k_{1}}{\varphi}\sin^{k_{2}}{\varphi}       
  & {\rm Evans}
    \end{array}
\end{equation}

\end{enumerate}

We have dealt with the curved SW and Evans systems as  natural generalizations of
central Hamiltonians, allowing for separation in spherical coordinates.
The symmetries, eigenfunctions and eigenvalues constitute a slight modification
of the curved KC and HO systems. Thus, the superintegrability character, as
well as the order of this kind of symmetries are easily checked in the frame
of the factorization method.

We should mention that there is another way to determine the symmetries of curved  HO and SW,  also based on factorizations, which is related to the curved Demkov-Fradkin tensor.
However, this question will be considered in a future work.

\section{Classical Analysis}

In this section we are going to consider the application of the factorization method in  canonical coordinates $(r,p_r,\theta,p_\theta,\varphi,p_\varphi)$. The general procedure is described in \cite{nuestra}, hence we will restrict ourselves to the application of this method  to the classical version of the quantum systems shown prior.

\subsection{ Curved Kepler-Coulomb}
Firstly, we will consider the classical analysis of the curved KC system. We will write the classical functions relevant to make up a symmetry of the system. 
%with associated ladder properties with the total angular momentum function. 
\begin{itemize}
    \item[(i)] {\bf Shift functions.} The factorization of the reduced Hamilonian:
    \begin{equation}
     \mathcal{H}_{\ell}= p^2_r+\frac{\ell^2}{\mathrm{S}^2_{\kappa}(r)}-\frac{q}{\mathrm{T}_{\kappa}(r)}=E
\end{equation}
leads to one pair of shift functions: 
\begin{equation}\label{sr}
{}_{r}\Sigma_{\ell}^{\pm}=
\mp\, i\, p_r+\frac{\ell}{\mathrm{S}_{\kappa}(r)}-\frac{q}{2\ell}\,, \, \qquad{}_{r}\Sigma_{\ell}^+{}_{r}\Sigma_{\ell}^-= \mathcal{H}_{\ell}+\frac{q^2}{4 \ell^2}
%\qquad
%\spc{r \lpm{r}}\smc{r \lpm{r}}
\end{equation}
Whose PB with the Hamiltonian are curvature independent and have shift properties associated to the total angular momentum $\ell$ :
\begin{equation}
     \{ \mathcal{H}_{\ell}, {}_{r}\Sigma_{\ell}^{\pm}\}_{L^2\to \ell^2} = \mp i \frac{2\ell}{\mathrm{S}^2_{\kappa}(r)}\, {}_{r}\Sigma_{\ell}^{\pm}=\mp  i \frac{\partial H_\ell }{\partial \ell}\, {}_{r}\Sigma_{\ell}^{\pm}
\end{equation}

    \item[(ii)]{\bf Symmetry functions.} The product of this radial shift functions with $\theta$ ladder functions, extracted from \cite{nuestra}, correspond to symmetries of the system:
    \begin{equation}
{}_{r,\theta}\mathcal{S}^\pm_{\ell}={}_{r}\Sigma^\pm_{\ell}{}_{\theta}\Lambda_{\ell}^{\pm};\quad \{H,{}_{r,\theta}\mathcal{S}^\pm_{\ell}\}_{L^2\to \ell^2}  = 0
    \end{equation}
    While acting as ladder functions of the total angular momentum:
    \begin{equation}
        \{\sqrt{L^2}\,, {}_{r\theta}\mathcal{S}^\pm_{\ell}\}  = \mp i\, {}_{r\theta}\mathcal{S}^\pm_{\ell}
    \end{equation}
    Due to their definition these symmetries are associated to the action-angle variables since  the phase of $\mathcal{S}^\pm_{\ell}$ functions corresponds to the associated coordinate of $\sqrt{L^2}$.
\end{itemize}
\subsection{Curved Harmonic Oscillator}
Next, we will will consider the curved HO system, 
\begin{itemize}
  \item[(i)]{\bf Shift functions.} The factorization of the reduced Hamiltonian:
    \begin{equation}\label{Htrigoclass}
 \mathcal{H}_{\ell}= p^2_r+\frac{\ell^2}{\mathrm{S}^2_{\kappa}(r)}+\frac{\Omega^2}{4}\mathrm{T}^2_{\kappa}(r)=E    
\end{equation} 
leads to two pair of independent functions:
\begin{equation}
\begin{array}{ll}
 \mathcal{H}_{\ell}={}_{r}a_\ell^+{}_{r} a_\ell^-+\ell(\kappa \ell-\Omega)\,,
& 
{}_{r}a_\ell^\pm=\mp i p_r+\frac{\ell}{\mathrm{T}_{\kappa}(r)}+\frac{\Omega}{2}\mathrm{T}_{\kappa}(r) \,,
%\quad &\{a_\ell^-,a_\ell^+\} =- i\omega
\\[2.ex]
 \mathcal{H}_{\ell}={}_{r}b^+_{\ell}   {}_{r}b_{\ell}^- +\ell(\kappa \ell+\Omega),\quad
 &
{}_{r}b_\ell^\pm=\mp i p_r+\frac{\ell}{\mathrm{T}_{\kappa}(r)}-\frac{\Omega}{2}\mathrm{T}_{\kappa}(r) \,,
%\quad & \{b_\ell^-,b_\ell^+\} = i\omega   
\end{array}
\end{equation}

Which can be coupled in order to obtain shift functions of the total angular momentum: \begin{equation}\label{shiftclasho}
    {}_{r}\Sigma_{\ell}^{\pm}= {}_{r}a_{\ell}^{\pm}{}_{r}b_{\ell}^{\pm}=- \mathcal{H}_{\ell}+ \frac{1+\mathrm{C}^2_{\kappa}(r)}{\mathrm{S}^2_{\kappa}(r)}\ell^2 \pm 2i \frac{\ell}{\mathrm{T}_{\kappa}(r)}p_{r}
\end{equation}
Where their PBs satisfy: 
\begin{equation}\label{shiftclashopb}
 \{ \mathcal{H}_{\ell}, {}_{r}\Sigma_{\ell}^{\pm}\}_{L^2\to \ell^2} = \mp i \frac{4\ell}{\mathrm{S}^2_{\kappa}(r)}\, {}_{r}\Sigma_{\ell}^{\pm}=\mp 2 i \frac{\partial H_\ell }{\partial \ell}\, {}_{r}\Sigma_{\ell}^{\pm}   
\end{equation}
While the opposite coupling leads to functions whose properties depend on the curvature value:
\begin{equation}
{}_{r}\Pi_{\ell}^{\pm}= {}_{r}b_{\ell}^{\pm}{}_{r}a_{\ell}^{\mp}; \quad \{ \mathcal{H}_{\ell}, {}_{r}\Pi^\pm_{\ell}\}_{L^2\to \ell^2} = \pm 2i\Omega \left(\kappa \mathrm{T}_{\kappa}(r)^2  +1\right)\, {}_{r}\Pi^\pm_{\ell}=\pm i \left( 4\kappa \frac{\partial  \mathcal{H}_{\ell} }{\partial \Omega}+2\Omega \right)\, {}_{r}\Pi^\pm_{\ell}
\end{equation}
They induce a energy change coupled with a frequency chance, in terms of the curvature value. Hence this functions act as energy ladder for the flat limit. 
        \item[(iii)]{\bf Symmetry functions.} The product of the radial shift functions and $\theta$ ladder functions $({}_{\theta}\Lambda_{\ell}^\pm)^2$, leads to symmetries of the system that act as ladder functions of the total angular momentum:
        \begin{equation}
            {}_{r,\theta}\mathcal{S}^\pm_{\ell}={}_{r}\Sigma_{\ell} ^\pm ({}_{\theta}\Lambda_{\ell}^\pm)^2;\quad \{ \mathcal{H}_{\ell}, {}_{r,\theta}\mathcal{S}^\pm_{\ell}\}_{L^2\to \ell^2}=0;\quad \{ \sqrt{L^2}, {}_{r,\theta}\mathcal{S}^\pm_{\ell}\}_{L^2\to \ell^2}=\pm 2i{}_{r,\theta}\mathcal{S}^\pm_{\ell}
        \end{equation}
            \item[(iv)]{\bf Ladder functions.} In order to determinate energy ladder functions we are going to consider a energy shift on the system:
             \begin{equation}\label{hclascos}
\overline{\mathcal{H}}_{\ell}= p^2_r+\frac{\ell^2}{\mathrm{S}^2_{\kappa}(r)}+\frac{\Omega^2}{4\kappa \mathrm{C}^2_{\kappa}(r)}=\overline{E}  ;   \quad   \overline{ \mathcal{H}_{\ell}}=   \mathcal{H}_{\ell} +\frac{\Omega^2}{4 \kappa}
\end{equation}
            The factorization of this Hamiltonian expression in terms of the energy leads two the following factorization functions:
        \begin{equation}
            \begin{array}{cc}
  {}_{r} c^\pm_{\overline{E}}=\mp i\kappa \mathrm{S}_{\kappa}(r)p_r+ \frac{\Omega}{2} \mathrm{C}_{\kappa}(r)+ \sqrt{\kappa \overline{E}}\mathrm{C}_{\kappa}(r)\,, \quad {}_{r}c^+_{\overline{E}}{}_{r}c^-_{\overline{E}}=  \kappa \overline{\mathcal{H}}_{\ell} +\frac{\Omega^2}{2}+\Omega \sqrt{\kappa \overline{E}}-\kappa^2 \ell^2  ,   \\
{}_{r}d^\pm_{\overline{E}}=\mp i\kappa \mathrm{S}_{\kappa}(r)p_r- \frac{\Omega}{2} \mathrm{C}_{\kappa}(r)+ \sqrt{\kappa \overline{E}}\mathrm{C}_{\kappa}(r)\,, \quad {}_{r}d^+_{\overline{E}}{}_{r}d^-_{\overline{E}}=  \kappa \overline{\mathcal{H}}_{\ell} +\frac{\Omega^2}{2}-\Omega \sqrt{\kappa \overline{E}}-\kappa^2 \ell^2  ,  & 
\end{array}
        \end{equation} 
Whose coupling corresponds to energy ladder functions of the curved HO system:
        \begin{equation}\label{ladderclasho}
{}_{r}\Lambda^{\pm}_{\overline{E}}=\frac{{}_{r}c^\pm_{\overline{E}}{}_{r}d^\pm_{\overline{E}}}{\kappa}=  \mathrm{C}^2_{\kappa}(2r)\overline{\mathcal{H}_{\ell}} +\kappa \ell^2-\frac{\Omega^2}{4\kappa}  \mp i \sqrt{\kappa \overline{E}}\mathrm{S}^2_{\kappa}(2r) p_{r};\quad \{ \overline{\mathcal{H}_{\ell}}, {}_{r}\Lambda^{\pm}_{\overline{E}}\}_{\overline{\mathcal{H}}_{\ell}\to \overline{E}}=\pm 4i \sqrt{\kappa \overline{E}}{}_{r}\Lambda^{\pm}_{\overline{E}}
        \end{equation}
\end{itemize}
From the phases of ${}_{r\theta}\mathcal{S}^\pm_{\ell}$ 
and ${}_{r}\Lambda^{\pm}_{\overline{E}}$ we are able to obtain the action-angle variables associated to $\sqrt{L^2}$ and the hamiltonian of the system, respectively.
\subsection{Smorodinsky-Winternitz and Evans Systems}
In this section we will consider the classical analysis of the SW and Evans systems (\ref{swep}).
\begin{itemize}
    \item[(i)]{\bf System identification.}  
These systems correspond to KC and HO  plus an additional centrifugal potential (\ref{swecp}), that modifies the angular momentum functions while maintaining the Hamiltonian structure:
\begin{equation}\label{reducedc}
\begin{array}{ll}
(a)\quad & L_z^2(\varphi,p_\varphi)   := p_\varphi^2+\frac{K_2^2}{\sin^2{\varphi}}+\frac{K_1^2}{\cos^2{\varphi}}=  m^2
\\[2.ex]
(b)\quad & {L}_m^2(\theta,p_\theta):=  p_\theta^2+\frac{m^2}{\sin^2{\theta}}+\frac{K_3^2}{\cos^2{\theta}}\,  
= \ell^2 
\\[2.ex]
(c)\quad & H_\ell(r,p_{r}): =  p_{r}^2 +V_r(r) +\frac{\ell^2}{r^2}    = E   
\end{array}
\end{equation}
This modified angular momentum functions can be identified as  classical P\"oschl-Teller Hamiltonians.
%, as they follow the structure:
%\begin{equation}\label{Hgen}
%  H(x,p_x)= p^2_x+\frac{a^2}{\sin^2{x}}+\frac{b^2}{\cos^2{x}}= c^2  
%\end{equation}
 \item[(ii)]{\bf Shift and ladder functions.} Due to this system identification, the shift and ladder functions of the angular problems will have the same structure as the ones associated to the curved HO, (\ref{shiftclasho}-\ref{ladderclasho}).
 %, taking into account the $a,b,c$ parameters (\ref{Hgen}) of each system. 
 
        \item[(iii)]{\bf Symmetry functions.} On one hand, all central systems with additional centrifugal potential (\ref{swecp}) will have a symmetry of the form: 
        \begin{equation}
            \mathcal{L}^\pm_{\theta,\varphi}={}_{\varphi}\Lambda^\pm_{m}{}_{\theta}\Sigma^\pm_{m};\quad 
\{L^2,\Lpmc{\theta,\varphi}\}_{L^2_z\to m^2}=\{H,\Lpmc{\theta,\varphi}\}_{L^2_z\to m^2}=0,\quad 
\{ \sqrt{L^2_z},\Lpmc{\theta,\varphi}\}_{L^2_z\to m^2}=\mp2 i\Lpmc{\theta,\varphi}
        \end{equation}
On the other hand, taking into account the KC and HO shift functions in \cite{nuestra} we can define the following symmetry of the system:
\begin{equation}
    S^\pm_{r,\theta}={}_{\theta}\Lambda^\pm_{\ell}{}_{r}\Sigma^\pm_{\ell};\quad 
\{L^2_z,S^\pm_{r,\theta}\}_{L^2\to \ell^2}=\{H,S^\pm_{r,\theta}\}_{L^2\to \ell^2}=0
\,,\qquad \ 
\{ \sqrt{L^2},S^\pm_{r,\theta}\}_{L^2\to \ell^2}=\mp2 iS^\pm_{r,\theta}
\end{equation}
\end{itemize}
Similarly, we can consider the analysis of the curved SW and Evans systems just substituting the radial shift functions of the curved HO and KC systems shown prior. 
%\vspace{5mm}

We are able to obtain action-angle variables of this problem from this symmetries. 

\section{Conclusions and Remarks}
Along this paper we have considered a generalization of the symmetry determination method  presented in \cite{nuestra}. First analyzing constant curvature systems \cite{ballesteros,perlick17},  where we have chosen to  the curved HO and KC systems in order to (i) compare the consistency with traditional methods \cite{carinena08,carinena12} as well as (ii) study the curvature effect on the resulting symmetries obtained by this method. Similarly, we considered the SW and Evans systems as non-central potentials systems examples \cite{winter,evans08,winternitz13}.
Some remarks surface from this analysis:
\begin{itemize}
\item Curved Analysis
\newline
 We have successfully implemented the factorization method to constant curvature systems considering a general curvature value $\kappa$. The quantum(classical) analysis of the HO and KC systems lead to curvature continuous symmetry operators(functions) that correspond to a generalization of the results obtained in \cite{nuestra}. Due to their definition, these operators allow us to determine the state spectrum of the system while their classical counterparts are closely related to the action-angle coordinates. 
    
    \item Identification Analysis
    \newline 
We considered the quantum(classical) analysis of the SW and Evans systems, since their potentials maintain the spherical separability of the system  (\ref{potang}) while being non-central, leading to modified angular momentum operators(functions). Where the application of the factorization method is immediate as we are able to consider a system identification. The $\varphi$ and $\theta$ effective systems correspond to one and two dimensional HO systems with curvature $\kappa=1$. Hence, we may apply the results obtained analyzing the curved HO system in order to solve these angular systems. This opens the factorization analysis to new superintegrable systems considering this identification with curved systems, imposing that they generate shift operators(functions), for the $\theta$ and $r$ problems, and ladder operators(functions)
for the $\varphi$ and $\theta$ problems.
\end{itemize}

Our aim in the near future is (i) to carry out a deep analysis of the Denkov-Fradkin tensor in non-euclidean spaces, as well as analyzing the period effect on the angular momentum operators \cite{kalnins12,gonera12,post,kalnins11,ttw09,ttw10}, considering the application of the factorization method to other coordinate systems \cite{evans90}, (ii) to consider non-constant curvature systems \cite{ballesteros2024dunkl} as well as (iii)  to generalize the factorization method in order to relativistic systems with spin.

\section{Acknowledgments}
The research of  Sergio Salamanca was supported by the European Union.-Next Generation UE/MICIU/Plan de Recuperacion, Transformacion y Resiliencia/Junta de Castilla y Leon.

\section{Appendix}
This section contains an extended version of the quantum analysis of the HO and SW-Evans sytems.
\subsection{Curved HO}
 In this section we are going to illustrate the quantum analysis of the curved HO system.
 \newline
We are working with the following radial state equation (\ref{hradial}):
\begin{equation}{\label{hradialtan2}}
\mathcal{H}_{\ell,\omega}R_n^{\ell}(r)=\left(-\partial_{rr}-\frac{2}{\mathrm{T}_{\kappa}(r)}\partial_{r}+\frac{\ell(\ell+1)}{\mathrm{S}^2_{\kappa}(r)}+\frac{\omega^2-\kappa^2}{4}\mathrm{T}^2_{\kappa}(r)\right) R_n^{\ell}(r)=E_{n}R_n^{\ell}(r)
        \end{equation}
    Considering a energy shift, we can express the prior problem in terms of the radial potential, $\overline{\mathcal{V_{HO}}}$:
\begin{equation}
     \mathcal{V_{HO}}=\frac{\omega^2-k^2}{4}\mathrm{T}^2_{\kappa}(r)=\frac{\omega^2-\kappa^2}{4\kappa\mathrm{C}^2_\kappa(r)}-\frac{\omega^2-\kappa^2}{4\kappa}=\overline{\mathcal{V_{HO}}}-\frac{\omega^2-\kappa^2}{4\kappa}
\end{equation} 
This new hamiltonian represents the same physical problem with a energy origin shift
$\overline{E_{n}}=E_{n}+\frac{\omega^2-\kappa^2}{4k}$:
\begin{equation}{\label{hradialcos2}}
\overline{\mathcal{H}}_{\ell,\omega}R_n^{\ell}(r)=\left(-\partial_{rr}-\frac{2}{\mathrm{T}_{\kappa}(r)}\partial_{r}+\frac{\ell(\ell+1)}{\mathrm{S}^2_{\kappa}(r)}+\frac{\omega^2-\kappa^2}{4k\mathrm{C}^2_{\kappa}(r)}\right) R_n^{\ell}(r)=\overline{E_{n}}R_n^{\ell}(r)
\end{equation}
Even though the hamiltonian $\overline{\mathcal{H}}_{\ell,\omega}$ is not continuous in the curvature, this divergence is removed when the energy shift is considered. The factorization of both hamiltonians lead to the same operators:
\begin{equation}\label{factHo2}
\begin{array}{ll}
\mathcal{H}_{\ell,\omega}=a^+_{\ell,\omega} a^-_{\ell,\omega}+\kappa(\ell(\ell+1)-\frac{1}{2})-\omega(\ell-\frac{1}{2})
%=a^-_{\ell+1} a^+_{\ell+1}-\frac{\omega}{2}(2\ell+3) 
& 
\left\{ \begin{array}{ll}
 a^+_{\ell, \omega}=-\partial_r+\frac{\ell-1}{\mathrm{T}_{\kappa}(r)}+\frac{\kappa+ \omega}{2}\mathrm{T}_{\kappa}(r) 
   
     &  \\
     a^-_{\ell, \omega}=\partial_r+\frac{\ell+1}{\mathrm{T}_{\kappa}(r)}+\frac{\kappa+ \omega}{2}\mathrm{T}_{\kappa}(r)
\end{array}\right.
 \\[2.ex]  
 \\[2.ex]
\mathcal{H}_{\ell,\omega}=b^-_{\ell+1,\omega+2\kappa} b^+_{\ell+1,\omega+2\kappa} +E_{l}=\left(\kappa(\ell(\ell+3)+\frac{3}{2})+\omega(\ell+\frac{3}{2})\right)\quad
%=
%b^+_{\ell} b^-_{\ell} +\frac{\omega}{2}(2\ell-1)  
&\left\{ \begin{array}{ll}
     b^+_{\ell, \omega}=-\partial_r+\frac{\ell-1}{\mathrm{T}_{\kappa}(r)}+\frac{\kappa- \omega}{2}\mathrm{T}_{\kappa}(r)  \\[2.ex]
  b^-_{\ell, \omega}=\partial_r=\frac{\ell+1}{\mathrm{T_{\kappa}(r)}}+\frac{\kappa-\omega}{2}\mathrm{T}_{\kappa}(r)   \\
\end{array}\right.
 \\[1.ex]  
 \end{array}
\end{equation}
In this case the factorization of the hamiltonian leads to two factorization pairs, connected by the inversion of $\omega$. While the states annihilated by each factorization operator satisfy the hamiltonian equation, only the states annihilated by $b^+$ can be normalized:
\begin{equation}
b^+_{n+1,\omega+2\kappa}R_{n}^{n}=0\rightarrow R_{n}^{n}\propto \mathrm{C}^{\frac{\kappa+ \omega}{2\kappa}}_{\kappa}(r)\mathrm{S}^{n}_{\kappa}(r);\quad \overline{\mathcal{H}}_{\ell,\omega}R_{n}^{n}=\overline{E}_{n}R_{n}^{n} ;\quad \overline{E}_{n}=\frac{\epsilon_{n}(\epsilon_{n}-4\kappa)}{4\kappa};\quad \epsilon_{n}=(\kappa(2n+5)+ \omega)
\end{equation}
Similarly to the KC problem, imposing the normalization condition onto $R_{n}^{n}$ states leads to a inequality condition on $n$ values. This inequality is obtained as well by imposing increasing energy values:
\begin{equation}
   \frac{\partial E_n}{\partial n}=k (3 + 2 n) + \omega>0
\end{equation}
While both $\overline{\mathcal{H}}_{\ell,\omega}$ and $\mathcal{H}_{\ell,\omega}$ systems have the same state spectra and are factored by the same operators $a^\pm, b^\pm$, the properties of these operators on both hamiltonians differ.
\vspace{5mm}

On one hand $a^\pm$ $b^\pm$ act as shift operators for both parameters on 
$\overline{\mathcal{H}}_{\ell,\omega}$:
\begin{equation}\label{commutover2}
    \begin{array}{cc}
a^+_{\ell+1,\omega-2\kappa} \overline{\mathcal{H}}_{\ell,\omega}=\overline{\mathcal{H}}_{\ell+1,\omega-2k}a^+_{\ell+1,\omega-2\kappa}, \qquad
  a^-_{\ell,\omega} \overline{\mathcal{H}}_{\ell,\omega}=\overline{\mathcal{H}}_{\ell-1,\omega+2\kappa}a^-_{\ell,\omega}
   \\[2.ex]
     b^+_{\ell+1,\omega+2\kappa}\overline{\mathcal{H}}_{\ell,\omega}=\overline{\mathcal{H}}_{\ell+1,\omega+2\kappa}b^+_{\ell+1,\omega+2\kappa}, \qquad
  b^-_{\ell,\omega} \overline{\mathcal{H}}_{\ell,\omega}=\overline{\mathcal{H}}_{\ell-1,\omega-2k}b^-_{\ell,\omega}
\end{array}
\end{equation}
On the other hand, since the energy shift between both Hamiltonians depends on the frequency value,
\[
\frac{\omega^2-\kappa^2}{4\kappa}\rightarrow \frac{(\omega\pm 2 \kappa)^2-\kappa^2}{4\kappa}=\frac{\omega^2-\kappa^2}{4\kappa}+(\kappa\pm \omega)
\] 
the factorization operators induce a energy shift on $\mathcal{H}_{\ell,\omega}$:
\begin{equation}\label{commut2}
    \begin{array}{cc}
a^+_{\ell+1,\omega-2\kappa} \mathcal{H}_{\ell,\omega}=\left(\mathcal{H}_{\ell+1,\omega-2k}+(\kappa-\omega)\right)a^+_{\ell+1,\omega-2\kappa}, \qquad
  a^-_{\ell,\omega} \mathcal{H}_{\ell,\omega}=\left(\mathcal{H}_{\ell-1,\omega+2k}+(\kappa+\omega)\right)a^-_{\ell,\omega}
   \\[2.ex]
     b^+_{\ell+1,\omega+2\kappa}\mathcal{H}_{\ell,\omega}=\left(\mathcal{H}_{\ell+1,\omega+2k}+(\kappa+\omega)\right)b^+_{\ell+1,\omega+2\kappa}, \qquad
  b^-_{\ell,\omega} \mathcal{H}_{\ell,\omega}=\left(\mathcal{H}_{\ell-1,\omega-2k}+(\kappa-\omega)\right)b^-_{\ell,\omega}
\end{array}
\end{equation}
We illustrate the effects of these operators in the following graph:
\begin{figure}[h]
	\centering
\includegraphics[width=7 cm]{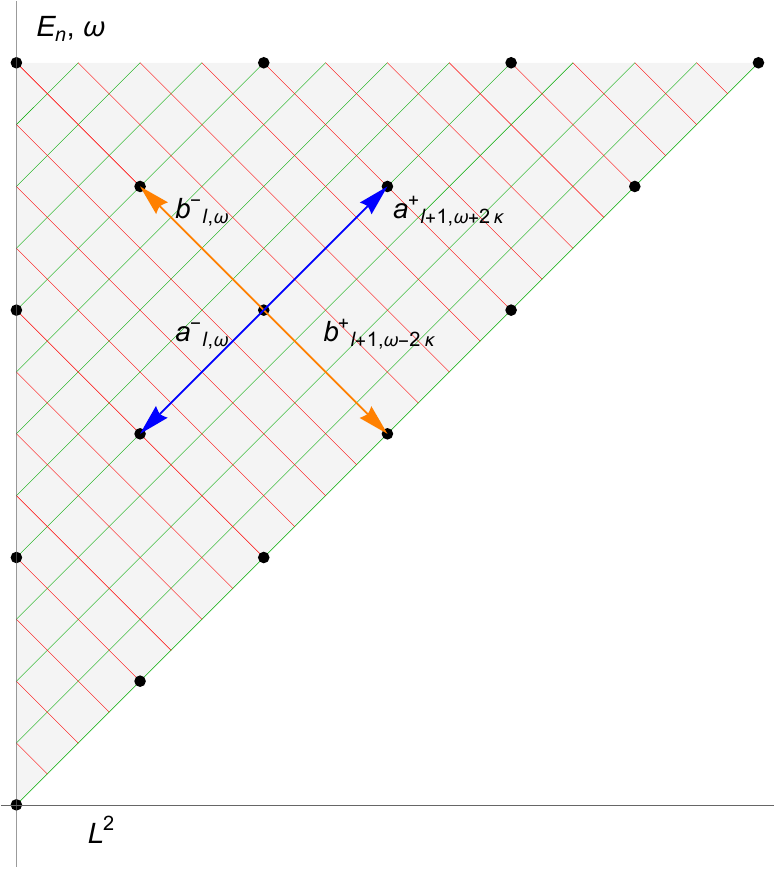}
\caption{Schematic representation of the action of the auxiliary operators $a_{\ell,\omega}^\pm(r)$ and $b_{\ell,\omega}^\pm(r)$ where each dot $(\ell,n,\omega)$ represent a state  $R_n^\ell(r)$ of the system $\mathcal{H}_{\ell,\omega}$.\label{figx}}
\end{figure}
\vspace{5mm}

Even though the factorization operators connect states of curved HO systems with different frequency values, we can define a shift operator that modifies the parameter $\ell$ while maintaining the frequency of the system:
\begin{equation}\label{shiftl2}
 \Sigma_{\ell}^+= a^+_{\ell+2,\omega}b^+_{\ell+1,\omega+2k}; \quad
   \Sigma_{\ell}^-= b^-_{\ell-1,\omega+2k}a^-_{\ell,\omega}  ; \quad  \Sigma_{\ell}^\pm \mathcal{H}_{\ell,\omega}= \mathcal{H}_{\ell\pm 2,\omega} \Sigma_{\ell}^\pm 
   \end{equation}

   They correspond to shift operators associated to the total angular momentum and allow us to generate the spectrum of state with a fixed energy value:
\begin{equation}\label{estadosradialesho2}
     \Sigma_{n}^+ R_{n}^{n}=0\rightarrow  R_{n}^{n}\propto \mathrm{C}^{\frac{\kappa+ \omega}{2\kappa}}_{\kappa}(r)\mathrm{S}^{n}_{\kappa}(r);\quad \Sigma_{\ell}^\pm R_{n}^{\ell} \propto R_{n}^{\ell\pm 2};\quad \overline{E}_{n}=\frac{\epsilon_{n}(\epsilon_{n}-4\kappa)}{4\kappa};\quad \epsilon_{n}=(\kappa(2n+5)+ \omega)
\end{equation} 
By coupling this radial shift operator with a $\theta$ ladder operator we are able to define the fifth independent symmetry of the system:
\[
{}_{r,\theta}\mathcal{S}^\pm_{\ell}={}_{r}\Sigma_{\ell} ^\pm {}_{\theta}\Lambda_{\ell\pm 1}^\pm{}_{\theta}\Lambda_{\ell}^\pm;  \quad {}_{r,\theta}\mathcal{S}^\pm_{\ell}\Psi_{n,\ell,m}(r,\theta,\varphi)\propto\Psi_{n,\ell\pm 2,m}
\]
This symmetry functions are curvature continuous and correspond to a generalization of the HO symmetry functions defined in \cite{nuestra}. 
While $a^\pm$ and $b^\pm$ functions correspond to a generalization of the factorization operator of the flat HO.  However are not able to decouple the frequency and energy change for the curved HO in terms of the auxiliary functions
since they induce a frequency change .
\vspace{3mm}

In order to define energy ladder operators we need to redefine the state equation, (\ref{hradialcos2}), in such a way that the new operator contains the energy as a parameter and the angular momentum  becomes the eigenvalue of the equation. Thus we define the operator  $\mathrm{N}_{n,\omega}$:
\begin{equation}\label{opintermedio2}
\mathrm{N}_{\epsilon,\omega}=\kappa^2\mathrm{S}^2_\kappa(r) (\overline{\mathcal{H}}_{\ell,\omega}-\overline{E}_{n})-\kappa^2\ell(\ell+1)=\kappa \left(-\kappa \mathrm{S}^2_\kappa(r)\partial_{rr}-\kappa\frac{2\mathrm{S}^2_\kappa(r)}{\mathrm{T}
_\kappa(r)}\partial_{r}-\frac{\epsilon(\epsilon-4k)}{4}\mathrm{S}^2_\kappa(r)+\frac{\omega^2-\kappa^2}{4} \mathrm{T}^2_\kappa(r) \right) 
            \end{equation}
Whose factorisation leads to the following functions:
\begin{equation}\label{nfact2}
\begin{array}{ll}
\mathcal{N}_{\epsilon,\omega}=c^+_{\epsilon,\omega} c^-_{\epsilon,\omega}+\frac{\kappa^2-(\epsilon+\omega)^2}{4}
%=a^-_{\ell+1} a^+_{\ell+1}-\frac{\omega}{2}(2\ell+3) 
& 
\left\{ \begin{array}{ll}
 c^+_{\epsilon, \omega}=-\kappa\mathrm{S}_{\kappa}(r)\partial_r+\frac{\kappa+\omega}{2\mathrm{C}_{\kappa}(r)}+\frac{\epsilon-2\kappa}{2}\mathrm{C}_{\kappa}(r) 
   
     &   \\[1.5ex]
      c^-_{\epsilon, \omega}=\kappa\mathrm{S}_{\kappa}(r)\partial_r+\frac{\kappa+\omega}{2\mathrm{C}_{\kappa}(r)}+\frac{\epsilon}{2}\mathrm{C}_{\kappa}(r) 
\end{array}\right.
 \\[2.ex]  
 \\[2.ex]
\mathcal{N}_{\epsilon,\omega}=d^+_{\epsilon,\omega} d^-_{\epsilon,\omega} +\frac{\kappa^2-(n-\omega)^2}{4}\quad
%=
%b^+_{\ell} b^-_{\ell} +\frac{\omega}{2}(2\ell-1)  
&
\left\{ \begin{array}{ll}
d^+_{\epsilon, \omega}=-\kappa\mathrm{S}_{\kappa}(r)\partial_r+\frac{\kappa-\omega}{2\mathrm{C}_{\kappa}(r)}+\frac{\epsilon-2\kappa}{2}\mathrm{C}_{\kappa}(r) 
   
     &   \\[1.5ex]
      d^-_{\epsilon, \omega}=\kappa\mathrm{S}_{\kappa}(r)\partial_r+\frac{\kappa-\omega}{2\mathrm{C}_{\kappa}(r)}+\frac{\epsilon}{2}\mathrm{C}_{\kappa}(r) 

\end{array}\right.
 \\[1.ex]  
 \end{array}
\end{equation}
%Since the four factorization operators will have the same flat limit. 
%\newline
The $c^\pm$ and $d^\pm$ operators have properties on  $\mathcal{N}_{\epsilon,\omega}$ analogous as the $a^\pm$ $b^\pm$ on $\overline{\mathcal{H}}_{\ell,\omega}$ (\ref{commutover2}) since they modify both parameters of $\mathcal{N}_{\epsilon,\omega}$:
\begin{equation}\label{ladderengk2}
    \begin{array}{cc}
c^+_{\epsilon-2\kappa,\omega-2\kappa} \mathcal{N}_{\epsilon,\omega}=\mathcal{N}_{\epsilon-2\kappa,\omega-2\kappa}c^+_{\epsilon-2\kappa,\omega-2\kappa}, \qquad
  c^-_{\epsilon,\omega} \mathcal{N}_{\epsilon,\omega}=\mathcal{N}_{\epsilon+2\kappa,\omega+2\kappa}c^-_{\epsilon,\omega}
   \\[2.ex]
    d^+_{\epsilon-2\kappa,\omega+2\kappa} \mathcal{N}_{\epsilon,\omega}=\mathcal{N}_{\epsilon-2\kappa,\omega+2\kappa}d^+_{\epsilon-2\kappa,\omega+2\kappa}, \qquad
  d^-_{\epsilon,\omega} \mathcal{N}_{\epsilon,\omega}=\mathcal{N}_{\epsilon+2\kappa,\omega-2\kappa}d^-_{\epsilon,\omega}
\end{array}
\end{equation}
%\newpage

This parameter change is described in the following graph:
\begin{figure}[h]
	\centering
\includegraphics[width=7 cm]{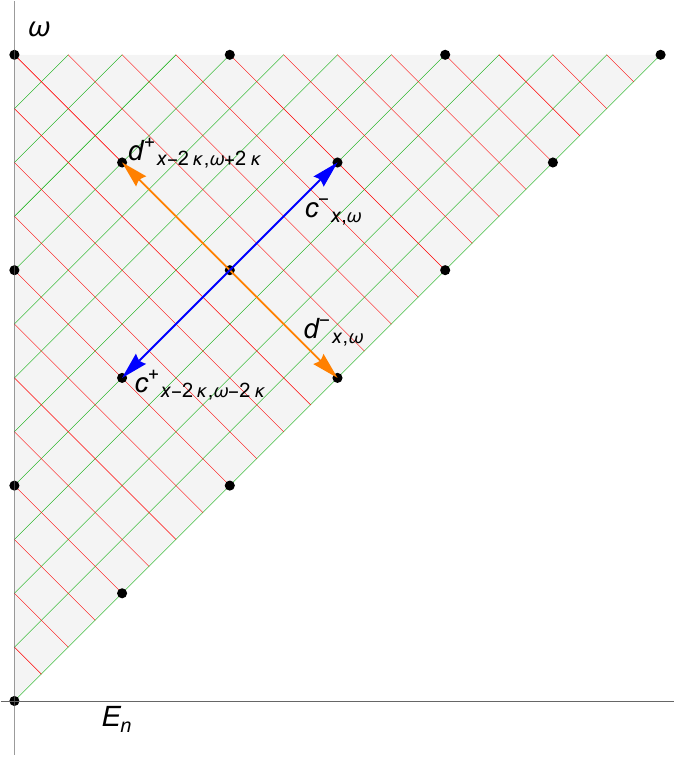}
\end{figure}

\vspace{5mm}
Similarly, we are able to isolate the energy change, leading to energy ladder operators:
\begin{equation}\label{ladderl2}
{}_{r}\Lambda_{\epsilon}^-=\frac{1}{\kappa}d^+_{\epsilon-4\kappa,\omega}c^+_{\epsilon-2\kappa,\omega-2\kappa};\quad _{r}\Lambda_{\epsilon}^-=\frac{1}{\kappa}d^-_{\epsilon+2 \kappa,\omega+2\kappa}c^-_{n,\omega}; \quad {}_{r}\Lambda_{\epsilon}^\pm \mathcal{N}_{\epsilon,\omega}=\mathcal{N}_{\epsilon\pm 4\kappa,\omega}{}_{r}\Lambda_{\epsilon}^\pm 
\end{equation}
Where we can recover the n parameter, taking into account the $\epsilon_{n}$ definition (\ref{estadosradialesho2}). We are able to define radial states with a fixed angular momentum from the effect of these operators:
\begin{equation}    
{}_{r}\Lambda_{n}^\pm \mathcal{N}_{n,\omega}=\mathcal{N}_{n\pm 2,\omega}{}_{r}\Lambda_{n}^\pm ,\qquad
{} _{r}\Lambda_{n}^-R_{n}^{n}=0\rightarrow R_{n}^{n}\propto \mathrm{C}^{\frac{\kappa+ \omega}{2\kappa}}_{\kappa}(r)\mathrm{S}^{n}_{\kappa}(r)  ; \quad _{r}\Lambda_{n}^\pm R_{n}^{\ell}\propto R_{n\pm 2}^{\ell}
\end{equation}
This ladder operators are continuous in the curvature and correspond to a curvature generalization of the flat energy ladder operators obtained in \cite{nuestra}.
We are able to modify each parameter of the system independently by the effect of the operators shown:
\begin{equation}\label{estadosho2}
{}_{r}\Lambda_{n}^\pm\Psi_{n,\ell,m}\propto\Psi_{n \pm 2,\ell,m};\quad {}_{r,\theta}\mathcal{S}^\pm_{\ell}\Psi_{n,\ell,m}\propto\Psi_{n,\ell\pm 2,m};\quad 
 L^\pm\Psi_{n,\ell,m}(r,\theta,\varphi)\propto\Psi_{n,\ell,m\pm 1}
\end{equation}
These operators allow us to travel though state solutions of the system as well as  obtain the state and eigenvalue spectrum, considering their creation and annihilation properties. However, since the operators ${}_{r}\Lambda_{n}^\pm$ and ${}_{r,\theta}\mathcal{S}^\pm_{\ell}$ modify the energy and total angular momentum of the states at a rate of two, states with odd and even energy values will have different definitions:
\begin{equation}
    \begin{array}{cc}
  \Psi_{n,\ell,m}\propto\left( (L^+)^m \prod_{j=0}^{y}{}_{r,\theta}\mathcal{S}^+_{2j} \prod_{i=0}^{q} {}_{r}\Lambda_{2i}^+\right)  \left( \Psi_{0,0,0}=\mathrm{C}^{\frac{\kappa+ \omega}{2\kappa}}_{\kappa}(r)\right)\quad n=2q: \ell=2y; \quad q,y\in \mathbb{N}   \\[2.5ex]
        \Psi_{n,\ell,m}\propto  \left( (L^+)^m \prod_{j=0}^{y}{}_{r,\theta}\mathcal{S}^+_{2j+1} \prod_{i=0}^{q} {}_{r}\Lambda_{2i+1}^+\right) \left( \Psi_{1,1,0}=\mathrm{C}^{\frac{\kappa+ \omega}{2\kappa}}_{\kappa}(r)\mathrm{S}_{\kappa}(r)\cos{\theta} \right) \quad n=2q+1: \ell=2y+1; \quad q,y\in \mathbb{N} 
    \end{array}
\end{equation}
While mantaining he parameter restrictions:
\begin{equation}\label{parametroskc2}
   0\leq |m| \leq \ell \leq n  ;\quad k (3 + 2 n) + \omega>0 ;\quad m,\ell,n \in \mathbb{N} 
\end{equation}

\subsection{Smorodinsky-Winternitz and Evans Systems}
In this section we are going to consider the analysis of the Smorodinsky-Winternitz and Evans systems in terms of the factorization method.  
\vspace{3mm}

This systems potentials can be interpreted as central potential systems, HO and KC respectively, 

\begin{equation}\label{swep2}
    V_{SW}=\frac{\Omega^2}{4}(x^2+y^2+z^2)+V_{c};\quad     V_{E}=-\frac{q}{\sqrt{x^2+y^2+z^2}}+V_{c}
\end{equation}
plus an additional centrifugal potential of the form:
\begin{equation}\label{swecp2}
    V_{c}=\frac{K_{1}^2}{x^2}+\frac{K_{2}^2}{y^2}+\frac{K_{3}^2}{z^2}=\frac{1}{r^2}\left(\frac{k_{1}(k_{1}-1)}{\cos^2{\theta}}+\frac{1}{\sin^2{\theta}}\left(\frac{k_{2}(k_{2}-1)}{\sin{\varphi}^2}+\frac{k_{1}(k_{1}-1)}{\cos{\varphi}^2}\right)\right); \quad k_{i}>0
\end{equation}
Leading to a total potential that follows the separability of the problem and can be interpreted as three one-dimensional potentials: 
\begin{equation}\label{potang2}
V(\varphi,\theta,r)=V_r(r)+ \frac{1}{r^2}\left(+V_{\theta}(\theta)+\frac{V_{\varphi}(\varphi)}{\sin^2{\theta}}\right)
\end{equation}
This potential structure allow us to separate the original three-dimensional problem considering states (\ref{phi}) into three one-dimensional connected ones:
\begin{equation}\label{reduced2}
\begin{array}{ll}
(a)\quad & \mathcal{{ L}}_z^2(\varphi) \phi_m(\varphi) := \left(-\partial_{\varphi\varphi}+\frac{k_2(k_2-1)}{\sin{\varphi}^2}+\frac{k_1(k_1-1)}{\cos{\varphi}^2}\right)\phi_m(\varphi)= m^2 \phi_m(\varphi)
\\[2.ex]
(b)\quad &  {\mathcal{{ L}}}_m^2(\theta)P_\ell^m(\theta):= \left(-\partial_{\theta\theta}-\frac{1}{\tan{\theta}}\partial_\theta+\frac{m^2}{\sin^2{\theta}}+\frac{k_3(k_3-1)}{\cos{\theta}^2}\, \right)
P_\ell^m(\theta) 
= \ell(\ell+1)\, P_\ell^m(\theta)
\\[2.ex]
(c)\quad & \mathcal{ H}_\ell(r)R_n^{\ell}(r): = \left(-\partial_{rr}-\frac{2}{r}\partial_r +V_r(r) +\frac{ \ell(\ell+1)}{r^2}\right) R_n^{\ell}(r) = E_n \, R_n^{\ell}(r)
\end{array}
\end{equation}
On the one hand, the addition of the centrifugal terms does not affect the radial problem, therefore we can recover the results of the central potential analysis of  the HO and KC radial problems, in particular  the shift and ladder radial operators,described in \cite{nuestra} as well as their curved version shown is present in this article.

\vspace{4mm}

On the other hand, the centrifugal potential leads to modified angular momentum  operators ${\mathcal{{ L}}}_m^2(\theta)$ and ${\mathcal{{ L}}}_z^2(\varphi)$ that can be identified as curved harmonic oscillators (\ref{hradialtan2}) since their potentials have same structure structure:
\begin{equation}\label{poteq2}
V=\frac{q(q+d-2)}{\sin^2{x}}+\frac{k_x(k_x-1)}{\cos^2{x}};\quad   q \rightarrow  k_{2},m,\ell  ;\quad k_{x} \rightarrow k_{1},k_{3}, \omega ; \qquad x\rightarrow \varphi,\theta,r
\end{equation}
Where d corresponds to the dimension of the system as $\{{\mathcal{{ L}}}_z^2(\varphi),{\mathcal{{ L}}}_m^2(\theta),\overline{\mathcal{H}}_{\ell,\omega}\}$ correspond to one, two and three dimensional systems respectively. 
We showcase the factorization of ${\mathcal{{ L}}}_m^2(\theta)$ in order to illustrate this identification:

%This identification translates to the factorization of ${\mathcal{{ L}}}_m^2(\theta)$ since the factorization operators follow the same strure as the ones in (\ref{factHo}):
\begin{equation}\label{facttheta2}
\begin{array}{ll}
{\mathcal{{ L}}}_m^2(\theta)= {}_{\theta}a^+_{k_{3},m}  {}_{\theta}a^-_{k_{3},m} +(k_3-m)(k_3-m+1)
%=a^-_{\ell+1} a^+_{\ell+1}-\frac{\omega}{2}(2\ell+3) 
& 
\left\{ \begin{array}{ll}
 {}_{\theta}a^+_{k_{3},m}=-\partial_{\theta}+\frac{m-1}{\tan{\theta}}+k_{3}\tan{\theta} 
   
     &  \\
    {}_{\theta}a^-_{k_{3},m}=\partial_{\theta}+\frac{m}{\tan{\theta}}+k_{3}\tan{\theta}
\end{array}\right.
 \\[2.ex]  
 \\[2.ex]
{\mathcal{{ L}}}_m^2(\theta)= {}_{\theta}b^+_{k_{3},m}  {}_{\theta}b^-_{k_{3},m} +(k_3+m-2)(k_3+m-1)\quad
%=
%b^+_{\ell} b^-_{\ell} +\frac{\omega}{2}(2\ell-1)  
&\left\{ \begin{array}{ll}
   {}_{\theta}b^+_{k_{3},m}=-\partial_{\theta}+\frac{m-1}{\tan{\theta}}-(k_{3}-1)\tan{\theta} 
   
     &  \\
    {}_{\theta}b^-_{k_{3},m}=\partial_{\theta}+\frac{m}{\tan{\theta}}-(k_{3}-1)\tan{\theta}
\end{array}\right.
 \\[1.ex]  
 \end{array}
\end{equation}

This equations follows the same structure as (\ref{factHo2}) and we can see explicitly the parameter relation described in (\ref{poteq2}).
\newline

This system identification allow us to obtain the eigenvalue structure as well as the allowed parameter values of $ \mathcal{{ L}}_z^2$ and ${\mathcal{{ L}}}_m^2$. While the eigenvalues of this systems maintain the structure of the free angular momentum operators $m^2$ and $\ell(\ell+1)$ respectively, the presence of the centrifugal potential modifies the allowed values of the $m$ and $\ell$ parameters. The centrifugal potential splits the space into octants as it diverges at $x,y,z=0$, this leads to a minimum value of the $m,\ell$ parameters as the angular momentum vector is locked in one octant:
\begin{equation}
  \ell\geq m+k_{3}\geq k_{1}+k_{2}+k_{3};\quad m=k_{1}+ k_{2}+ 2f; \quad \ell=k_{1}+k_{2}+k_{3}+2g;\quad f,g\in \mathbb{N} 
\end{equation}
While there are eight possible minimum values of this parameters, as the potential is invariant by the transformation $k_{i}\rightarrow 1-k_{i}$, since the angular momentum vector is locked in one octant we are able to select a value origin. 
\newline

We can take advantage of the results obtained from the analysis of the curved HO system, and define shift (\ref{shiftl2}) and ladder (\ref{ladderl2}) operators for the modified angular momentum operators from the parameter relations described in (\ref{poteq2}).

\begin{equation}
\begin{array}{cc}
   {}_{\varphi}\Lambda_{m}^\pm=\cos{2\varphi}L^2_z+(1\pm m)\sin{2\varphi}\partial_{\varphi}+k_{1}-{k_1^2}+k_{2}^2-k_2 \pm m \cos{2\varphi}  &  \\[2.ex]
   
{}_{\theta}\Sigma_{m}^\pm=-L^2-2\frac{1\pm m}{\tan{\theta}}\partial_{\theta}+k_{3}(k_{3}-1)-m(m\pm1)+\frac{2m(m\pm1)}{\sin^2{\theta}}
 & \\[2.ex]
 {}_{\theta}\Lambda_{\ell}^\pm=\cos{2\theta}L^2-(1\pm (\ell+\frac{1}{2}))\partial_{\theta}+L_z^2-k_{3}(k_{3}-1)+(\frac{1}{2}\pm(\ell+\frac{1}{2}))(2\cos^2{\theta}+\sin^2{\theta})
\end{array}
\end{equation}
 
These operators allow us to obtain the state and eigenvalue spectrum of the angular problems in terms of their creation and anhilation properties: 
\begin{equation}
    \begin{array}{cc}\label{estadosanglares2}
      {}_{\varphi}\Lambda_{k_{1}+k_{2}}^-   \phi_{k_{1}+k_{2}}=0\rightarrow \phi_{k_{1}+k_{2}}= \cos^{k_{1}}{\varphi}\sin^{k_{2}}{\varphi};\quad  {}_{\varphi}\Lambda_{m}^\pm   \phi_m \propto \phi_{m\pm2} ;\quad L^2_z\phi_m=m^2\phi_m &  \\[2.ex]
      {}_{\theta}\Sigma_{m}^+ P^m_{m+k_{3}}=0\rightarrow  P^m_{m+k_{3}}=\sin^{m}{\theta}\cos^{k_{3}}{\theta} ;\quad   {}_{\theta}\Sigma_{m}^+ P^m_{\ell}\propto P^{m\pm2}_{\ell}\quad {\mathcal{{ L}}}_m^2P^m_{\ell}=\ell(\ell+1)P^m_{\ell} & \\[2.ex]
       {}_{\theta}\Lambda_{m+k_{3}}^- P^m_{m+k_{3}}=0\rightarrow  P^m_{m+k_{3}}=\sin^{m}{\theta}\cos^{k_{3}}{\theta} ;\quad {}_{\theta}\Lambda_{\ell}^- P^{\ell}_{m}\propto P^{\ell\pm 2}_{m};\quad  {\mathcal{{ L}}}_m^2P^m_{\ell}=\ell(\ell+1)P^m_{\ell} & 
    \end{array}
\end{equation}
From this shift and ladder operators we can define  symmetry functions that act on the complete state of the system. While we can always define a symmetry operator associated to the parameter $m$, the existence of an operators associated to $\ell$ requires the existence of $ {}_{r}\Sigma_{\ell}^\pm$ operators and their rate of modification of the $\ell$ parameter. Since the associated radial potential of the SW and Evans systems corresponds to the KC and HO potentials, we can directly consider the results obtained for their curved version,  from  (\ref{kcurvo}),(\ref{shiftl2}) and from this shift operators consider the following symmetries:
\begin{equation}
     {}_{\theta,\varphi}\mathcal{L}^\pm_{m}={}_{\theta}\Sigma_{m} ^\pm {}_{\varphi}\Lambda_{m}^\pm \quad {}_{\theta,\varphi}\mathcal{L}^\pm_{m}\Psi_{n,\ell,m}\propto\Psi_{n,\ell,m\pm 2};\quad      {}_{r,\theta}\mathcal{S}^\pm_{\ell}={}_{r}\Sigma_{\ell} ^\pm {}_{\theta}\Lambda_{\ell}^\pm; \quad {}_{r,\theta}\mathcal{S}^\pm_{\ell}\Psi_{n,\ell,m}\propto\Psi_{n,\ell\pm 2,m}
\end{equation}
Considering the annihilation properties of this symmetries we have the following bound states: 
\begin{equation}
    \begin{array}{cc}
  \Psi_{\Sigma k_{i},\Sigma k_{i},k_{1}+k_{2}}=  \mathrm{C}^{\frac{\kappa+ \omega}{2\kappa}}_{\kappa}(r)
  \mathrm{S}^{\Sigma k_{i}}_{\kappa}(r)\left(\sin{\theta}\right )^{k_{1}+k_{2}} \cos^{k_{3}}{\theta}\cos^{k_{1}}{\varphi}\sin^{k_{2}}{\varphi} &  SW\\[1.5ex]
  \Psi_{\Sigma k_{i},\Sigma k_{i},k_{1}+k_{2}}= e^{\frac{-q r}{2(n+1)}} 
  \mathrm{S}^{\Sigma k_{i}}_{\kappa}(r)\left(\sin{\theta}\right )^{k_{1}+k_{2}} \cos^{k_{3}}{\theta}\cos^{k_{1}}{\varphi}\sin^{k_{2}}{\varphi}       & Evans
    \end{array}
\end{equation}
The presence of the centrifugal terms leads to minimum values of all parameters of the system:
\begin{equation}
m= k_{1}+k_{2}+2f;\quad \ell= m+k_{3}+2g;\quad n=\ell+2h;\quad f,g,h\in \mathbb{N}
\end{equation}


\begin{thebibliography}{}
\bibitem{nuestra}Kuru, Ş., Negro, J. \& Salamanca, S. Quantum, classical symmetries and action-angle variables by factorization of superintegrable systems. {\em ArXiv E-prints}. pp. arXiv-2305 (2023)

\bibitem{ballesteros}Herranz, F., Ballesteros, Á. \& Others Superintegrability on three-dimensional Riemannian and relativistic spaces of constant curvature. {\em SIGMA. Symmetry, Integrability And Geometry: Methods And Applications}. \textbf{2} pp. 010 (2006)
\bibitem{carinena08}Cariñena, J., Ranada, M. \& Santander, M. The harmonic oscillator on Riemannian and Lorentzian configuration spaces of constant curvature. {\em Journal Of Mathematical Physics}. \textbf{49} (2008)

\bibitem{carinena12}Carinena, J., Ranada, M. \& Santander, M. Curvature-dependent formalism, Schrödinger equation and energy levels for the harmonic oscillator on three-dimensional spherical and hyperbolic spaces. {\em Journal Of Physics A: Mathematical And Theoretical}. \textbf{45}, 265303 (2012)

\bibitem{ballesteros2014new}Ballesteros, A., Blasco, A., Herranz, F. \& Musso, F. A new integrable anisotropic oscillator on the two-dimensional sphere and the hyperbolic plane. {\em Journal Of Physics A: Mathematical And Theoretical}. \textbf{47}, 345204 (2014)

\bibitem{najafizade2021behavior}Najafizade, A. \& Panahi, H. Behavior of a constrained particle on superintegrability of the two-dimensional complex Cayley–Klein space and its thermodynamic properties. {\em Physica A: Statistical Mechanics And Its Applications}. \textbf{573} pp. 125935 (2021)
\bibitem{evans90}Evans, N. Superintegrability in classical mechanics. {\em Physical Review A}. \textbf{41}, 5666 (1990)

\bibitem{herranz2005maximally}Herranz, F., Ballesteros, A., Santander, M. \& Sanz-Gil, T. Maximally superintegrable Smorodinsky-Winternitz systems on the N-dimensional sphere and hyperbolic spaces. {\em ArXiv Preprint Math-ph/0501035}. (2005)

\bibitem{shmavonyan2019cn}Shmavonyan, H. CN-Smorodinsky–Winternitz system in a constant magnetic field. {\em Physics Letters A}. \textbf{383}, 1223-1228 (2019)

\bibitem{evans08}Verrier, P. \& Evans, N. A new superintegrable Hamiltonian. {\em Journal Of Mathematical Physics}. \textbf{49} (2008)

\bibitem{winternitz13}Miller, W., Post, S. \& Winternitz, P. Classical and quantum superintegrability with applications. {\em Journal Of Physics A: Mathematical And Theoretical}. \textbf{46}, 423001 (2013)

\bibitem{ttw09}Tremblay, F., Turbiner, A. \& Winternitz, P. An infinite family of solvable and integrable quantum systems on a plane. {\em Journal Of Physics A: Mathematical And Theoretical}. \textbf{42}, 242001 (2009)

\bibitem{ttw10}Tremblay, F., Turbiner, A. \& Winternitz, P. Periodic orbits for an infinite family of classical superintegrable systems. {\em Journal Of Physics A: Mathematical And Theoretical}. \textbf{43}, 015202 (2009)


\bibitem{enciso}Ballesteros, A., Enciso, A., Herranz, F. \& Ragnisco, O. Hamiltonian systems admitting a Runge–Lenz vector and an optimal extension of Bertrand’s theorem to curved manifolds. {\em Communications In Mathematical Physics}. \textbf{290} pp. 1033-1049 (2009)



\bibitem{winter}Post, S. \& Winternitz, P. An infinite family of superintegrable deformations of the Coulomb potential. {\em Journal Of Physics A: Mathematical And Theoretical}. \textbf{43}, 222001 (2010)
\bibitem{kalnins12}Kalnins, E. \& Miller Jr, W. Structure results for higher order symmetry algebras of 2D classical superintegrable systems. {\em ArXiv Preprint ArXiv:1101.5292}. (2011)
\bibitem{gonera12}Gonera, C. On the superintegrability of TTW model. {\em Physics Letters A}. \textbf{376}, 2341-2343 (2012)
\bibitem{post}Lévesque, D., Post, S. \& Winternitz, P. Infinite families of superintegrable systems separable in subgroup coordinates. {\em Journal Of Physics A: Mathematical And Theoretical}. \textbf{45}, 465204 (2012)

\bibitem{kalnins11}Kalnins, E. \& Miller Jr, W. Structure results for higher order symmetry algebras of 2D classical superintegrable systems. {\em ArXiv Preprint ArXiv:1101.5292}. (2011)


\bibitem{ballesteros2024dunkl}Ballesteros, A., Najafizade, A., Panahi, H., Hassanabadi, H. \& Dong, S. The Dunkl oscillator on a space of nonconstant curvature: An exactly solvable quantum model with reflections. {\em Annals Of Physics}. \textbf{460} pp. 169543 (2024)

\bibitem{perlick17}Kuru, Ş., Negro, J. \& Ragnisco, O. The Perlick system type I: From the algebra of symmetries to the geometry of the trajectories. {\em Physics Letters A}. \textbf{381}, 3355-3363 (2017)
\bibitem{kuru2020general}Kuru, Ş., Marquette, I. \& Negro, J. The general Racah algebra as the symmetry algebra of generic systems on pseudo-spheres. {\em Journal Of Physics A: Mathematical And Theoretical}. \textbf{53}, 405203 (2020)
\bibitem{panahi2017two}Panahi, H. \& Nemati, M. Two-Dimensional Exactly Solvable Quantum Model Obtained from SU (3)/SU (2) Homogenous Space. {\em International Journal Of Theoretical Physics}. \textbf{56} pp. 2265-2270 (2017)
	 

\end{thebibliography}
\end{document}